\documentclass[journal=aamick,manuscript=article]{achemso}

\usepackage[version=3]{mhchem} 
\usepackage[T1]{fontenc}       
\usepackage{color}
\usepackage{amsmath}
\usepackage{float}
\usepackage{txfonts}
\usepackage{enumitem} 					
\usepackage{hyperref}
\usepackage[labelfont=bf]{caption} 
\usepackage[geometry]{ifsym} 
\usepackage{color} 
\usepackage{wasysym}	
\definecolor{mycol1}{RGB}{255, 231, 0}
\definecolor{mycol2}{RGB}{255, 132, 0}
\definecolor{mycol3}{RGB}{255, 33, 0}
\definecolor{mycol4}{RGB}{140, 0, 0}
\definecolor{mycol5}{RGB}{132, 0, 255}
\definecolor{mycol6}{RGB}{214, 0, 0}
\definecolor{mycol7}{RGB}{0, 0, 255}

\usepackage{graphicx,calc}
\author{Slav A. Semerdzhiev}
\affiliation[Wageningen University]
{Physical Chemistry and Soft Matter, Wageningen University and Research, Stippeneng 4, 6708 WE Wageningen, The Netherlands}
\alsoaffiliation[DPI]
{Dutch Polymer Institute (DPI), P.O. Box 902, 5600 AX Eindhoven, The Netherlands}

\author{Hanne M. van der Kooij}
\affiliation[Wageningen University]
{Physical Chemistry and Soft Matter, Wageningen University and Research, Stippeneng 4, 6708 WE Wageningen, The Netherlands}
\alsoaffiliation[DPI]
{Dutch Polymer Institute (DPI), P.O. Box 902, 5600 AX Eindhoven, The Netherlands}

\author{Remco Fokkink}
\author{Joris Sprakel}

\email{joris.sprakel@wur.nl}
\affiliation[Wageningen University]
{Physical Chemistry and Soft Matter, Wageningen University and Research, Stippeneng 4, 6708 WE Wageningen, The Netherlands}

\title{Rapid and unambiguous determination of the open time for waterborne paint films with Laser Speckle Imaging.  }
\abbreviations{IR,NMR,UV}

\begin{document}

\begin{tocentry}

Some journals require a graphical entry for the Table of Contents.
This should be laid out ``print ready'' so that the sizing of the
text is correct.

Inside the \texttt{tocentry} environment, the font used is Helvetica
8\,pt, as required by \emph{Journal of the American Chemical
Society}.

The surrounding frame is 9\,cm by 3.5\,cm, which is the maximum
permitted for  \emph{Journal of the American Chemical Society}
graphical table of content entries. The box will not resize if the
content is too big: instead it will overflow the edge of the box.

This box and the associated title will always be printed on a
separate page at the end of the document.

\end{tocentry}

\begin{abstract}
In drying coatings, the time span between the deposition of the wet film and the formation of an uniform solid harbors a procession of complex phenomena each of which carves the final properties of the dry paint film. Attaining a better control of the drying process requires the availability of a tool that allows monitoring the attending dynamics. The experimental access to the dynamics underlying the drying process is challenging as paint films are often opaque and the phenomena accompanying the drying process usually populate a multitude of time and length scales. To circumvent these obstacles we deploy Laser Speckle Imaging (LSI) as a technique that allow us to quantitatively probe the internal dynamics of drying waterborne paints and unambiguously determine a crucial handling parameter such as the open time. We develop a set of scaling relations which accurately predict the behavior of the experimentally determined open time as a function of a parameter set that governs the drying process. Additionally, we harness the big temporal dynamic range of LSI to monitor phenomena that inhabit the later stages of the drying process. Remarkably, we not only capture the film formation time, the time point in which phase inversion takes place, but also microscopically detect the deformation of polymeric particles that precedes it.   
\end{abstract}

\newpage
\section{Introduction}

\par
The drying of  thin aqueous colloidal films has been for some time now a phenomenon that enjoys continuous scientific scrutiny. The entanglement of numerous accompanying processes renders the drying phenomenon rather complex and has harnessed  significant scientific efforts in a pursuit of a better understanding. The need of such has also a practical genesis.The drying of composite colloid films is of great importance for numerous industrial fields such as  food technology, inkjet printing and particularly in coatings technology.\cite{Embuscado2009, Talbot2015, Keddie2010} The usage of organic solvent based coatings dates long way back and that has provided enough time for the industry to continuously tailor this type of products to a level which ensures a satisfactory performance of the coating during the drying and post-drying period. However, while drying, solvent based coatings emit volatile organic compounds (VOCs) which are very potent in amplifying the green house effect and at the same time appear as hazardous compounds for the health of the consumer and the professional applying it.\cite{Wicks1992, Morrow1995, Triebig2001, Böckelmann2003} This has set the intention of legislators and manufacturers to take on a transition course from solvent based to VOCs free aqueous systems. This transition however is far from being trivial and offers number of challenges. The complex nature of the drying and solid film formation process in waterborne coatings impede the progress in understanding and controlling these phenomena. Also, the introduction of water as a continuous phase places 2 opposing requirements: the binder particles, which are typically hydrophobic, need to be colloidally stable  to avoid aggregation while the product is on the shelf, but not too stable since that would hinder the film formation in the final stages of the drying process. Meeting those demands requires smart chemical engineering strategies, and to generate such, a better understanding of the intricate drying processes is also imperative. There is another `inconvenient'  aspect of water being the continuous phase. As a solvent, water evaporates relatively faster compared to its organic counterparts. This significantly shortens the time window, a crucial parameter for coatings also known as the open time, in which the deposited film can be handled without leaving permanent topological defects.\par
The dynamics underlying the drying process determine the ultimate fate of a drying film of aqueous suspensions with soft particles, and thus its final properties. The first step towards solving the listed above challenges is establishing the right tool that would enable us to study and understand these drying dynamics. Unfortunately, that is a challenge in itself. Visualizing the dynamics could be notoriously difficult because they are often heterogeneous and span a wide range of time scales.  Additionally, often being highly turbid, paint films do not allow investigations relying on conventional techniques such as microscopy and dynamic lights scattering, which operate in the limit of sufficiently transparent samples. The lack of proper means to monitor and study film drying, obscures the progress in understating this process and makes it difficult to determine important quantities such as the open time. It is not surprising that existing methods for open time determination are rather qualitative and are performed manually from human operators which can introduce ambiguity and big spread in the obtained results. 
\par
In this work, we use Laser Speckle Imaging (LSI) to elucidate with microscopic resolution the dynamics taking place deep inside drying opaque latex films and spanning many decades of time. The access to these internal dynamics allow us to determine the open time for realistic paint systems in a quantitative and unambiguous manner, and for a wide range of conditions. We validate our LSI findings by systematically comparing the experimentally determined open time behavior as a function of different environmental parameters with  forecasts stemming from scaling relations based on Langmuir's theoretical framework about evaporation. Finally, we expand further the application scope of LSI by demonstrating the ability of this technique to follow in detail the solid film formation in the later stages of the drying process. Not only we can readily detect the onset of phase inversion in the paint film but can also track the behavior of this phenomenon as a function of an important parameter such as hardness of the latex particles.        
\newline

\section{Results and discussion}
\subsection{Laser speckle imaging and the open time of waterborne paint films.}
Laser speckle imaging (LSI) belongs to the family of techniques that exploits the
phenomenon of light scattering to monitor the internal dynamics of different types of
systems. In contrast to conventional dynamic light scattering (DLS), LSI operates in the
limit of optically thick samples. This characteristic feature has encouraged the
deployment of LSI in numerous fileds such as biology, food
technology, material and surface science.\cite{Zhou2006, Zakharov2009, Draijer2009, Pajuelo2003, Maksymenko2015, Hajjarian2012, Erpelding2008, Hinsch2000, Fricke-Begemann2004, Koshoji2015} In a medium that is densely populated with scattering species, light propagation is accompanied by multiple scattering
events. After certain number of such events the propagation direction of the photons is
completely randomized (Fig. \ref{fig1}a, inset). This allows to treat photon propagation in optically thick medium (on length scales that ensure sufficient number of scattering events) as a diffusive transport.\cite{Pine1990}  At a distance $r$ from the boundary of the sample - usually the location of the detector - the scattered photons superpose and generate a speckle pattern due to the trajectory lengths of the paths that they have undergone in the scattering medium. Since light is much faster than any type of molecular, or particle motion in the case of colloid suspension, the momentary realization of this interference pattern is directly related to the instantaneous arrangement of the
scattering species within the sample. At longer timescales the speckle pattern changes due
to the displacement of the scattering particles. LSI exploits the diffusive nature of light propagation in turbid media to extract information from these temporal changes in the speckle patterns about the movement of scattering species, and thus about the ongoing internal dynamics in highly scattering systems.\cite{Kooij2016}
To quantify the intensity fluctuations in a speckle pattern, we resort to the use of the
intensity structure factor $d_2$ which reads as:\cite{Schulz-DuBois1981, Schatzel1987}
\begin{equation}
\label{eq. 1}
d_2(t,x,y,\tau)=\frac{\langle[I(t,x,y)-I(t+\tau,x,y)]^2\rangle}{\langle I(t,x,y) \rangle \langle I(t+\tau,x,y)\rangle}
\end{equation}
Here $t$ is time, x and y are spatial coordinates and $\tau$ is the time lag. High and low $d_2$ values indicate high and low dynamics respectively. The $d_2$ is a convenient quantity allowing us to temporally and spatially map the dynamic activity in the system of interest. Additionally, we can selectively focus on dynamics populating different time scales by choosing the appropriate $\tau$ which sets the lag time between the speckle pattern snapshots we are trying to correlate (Fig. \ref{fig1}b). The ability to construct spatio - temporal $d_2$ maps allow us to monitor the evolution of the dynamics and easily detect any dynamic heterogeneity that eventually might occur within the sample under investigation. Detecting such heterogineities is crucial for the next steps in the analysis in which the $g_2$, the intensity autocorrelation function that is closely related to the $d_2$, needs to be computed and averaged over time and all the speckles in the field of view. The temporal intensity autocorrelation function is given by:
\begin{equation}
\label{eq. 2}
g_2(t,x,y,\tau)=\frac{\langle I(t,x,y)I(t+\tau,x,y)\rangle}{\langle I(t,x,y) \rangle \langle I(t+\tau,x,y)\rangle}
\end{equation}
Having the $g_2(t,\tau)$ calculated grants us access to the field correlation function $g_1(t,\tau)$ via the Siegert relation:
\begin{equation}
\label{eq. 3}
g_1(t,\tau)=\frac{1}{\sqrt{\beta}}\sqrt{g_2(t,\tau)-1}
\end{equation}
where $\beta$ is a numerical  prefactor depending on the experimental setup (see Experimental section).\cite{Pine1990} Finally, we can fit $g_1(t,\tau)$ using a single-exponential decay function:
\begin{equation}
\label{eq. 4}
g_1(t,\tau)=e^{-\gamma (\tau / \tau_0 (t))^{\alpha(t)}}
\end{equation}
where $\tau_0(t)$ is the characteristic relaxation time, $\alpha(t)$ is a stretching exponent and $\gamma$ is an experimental numerical constant that has been determined elsewhere (see Experimental section). Both, the relaxation time and the stretching exponent, could be viewed as means to extract more quantitative information about the ongoing dynamics. Since the relaxation time $\tau_0=1/{k_0}^2D$ is related to the diffusion rate of the scatterers, this quantity provides information about the resistance (effective viscosity) that the scatterers experience while they are moving. The stretching exponent  $\alpha(t)$ on the other hand signals the type of motion that the scattering particles undergo. Since $g_1(t,\tau)=e^{-\gamma (k_0 \sqrt{\langle \Delta x^2(\tau) \rangle})}$ where $\langle \Delta x^2(\tau) \rangle$ is the mean square displacement, values of $\alpha<0.5$ imply sub-diffusive dynamics ($\langle \Delta x^2(\tau)\rangle\sim\tau^{\alpha}$) while $\alpha\rightarrow1$ suggests for a ballistic motion ($\langle \Delta x^2(\tau) \rangle \sim \tau^2$) of the particles.\cite{Guo2012, Weitz} The Brownian dynamics ($\langle \Delta x^2(\tau)\rangle\sim\tau$) are recovered at $\alpha=0.5$. The sensitivity of both parameters, $\alpha$ and $\tau_0$, to changes in the motion of the scattering particles renders these two quantities as complementary tools appropriate for monitoring the dynamics in evolving systems such as drying waterborne paint films.\par

\par
The drying of thin colloidal films is complex phenomenon accompanied by a constellation of processes each of which is often marked by its own characteristic dynamics, time and length scale.\cite{Keddie1997,Kooij2016, Kooij2016b} The drying process of a film of soft colloids could be tentatively divided in four main stages (Fig. \ref{fig2}a). We start with a solution of freely diffusing particles. Due to solvent evaporation, in time the particles get concentrated and more densely packed. Then under the work of capillary forces the soft particles start to deform and at some point, if the setting is right (sufficiently low glass transition temperature of the soft polymer particles), the particles coalesce to give an uniform solid film. The big temporal dynamic range of the LSI technique allow us to capture different processes spanning multiple decades of time, and thus many of the events that take place during the drying of such films.\cite{Kooij2016}. We first focus on the early drying stages of waterborne paint systems. The dynamics populating this time scales govern the open time ($t_{OT}$) of the drying film which is a crucial handling parameter not only for coatings but for any type of system containing an evaporating solvent. By definition, open time is the period after applying the paint in which the film can still be reworked without leaving any visible and permanent topological irregularities. Since LSI provides access to the early dynamics in drying paint films, we can determine the open time in a quantitative and unambiguous manner by following them. To demonstrate that, we start by depositing a 200 $\mu$m thick film of an acrylic emulsion on a glass substrate and record a sequence of raw LSI images from the central section of the drying film. The emulsion contains TiO$_2$ nano particles which are the dominant scattering species, and hence, the reporters for of onging dyanamics in the deposited film. The $d_2$ maps reveal (not shown here) spatially uniform evolution of the dynamics in time which allow us to proceed towards the computation of the average field correlation curves $g_1(\tau)$ with equations \ref{eq. 2} and \ref{eq. 3} (Fig. \ref{fig2}b). In time, the $g_1(\tau)$ - curves shift towards longer correlation times $\tau$ which indicates retardation in the particle motion as a consequence of the increased solid content caused by the solvent evaporation. By examining the time evolution of the relaxation time $\tau_0$ and the stretching exponent $\alpha$ we can clearly discern different regimes in the dynamics. The $\tau_0(t)$ curve exhibits a sharp change in the slope $d\tau / dt$ (Fig. \ref{fig2}c). This characteristic kink is read as the transition from a slow to a faster increase in the viscosity of the paint film. Since viscosity is directly related to the flow properties of a substance, this time point could be interpreted as the onset of deterioration in the workability of the paint film. Interestingly, the $\alpha(\tau)$ curve shows similar sudden increase in the slope $d\alpha / dt$ (Fig. \ref{fig2}d). This kink however, is shifted to a later time point and has a different origin. The dynamics accompanying the period preceding this time point t < 1100 s are close to Brownian ($\alpha\sim$ 0.7) and barely change whereas the subsequent period (t > 1100 s) gets gradually dominated by advective processes ($\alpha\rightarrow$ 1) powered by the ongoing evaporative flux. The transition between these distinct regimes  marks the transformation of the liquid paint film into a quasi-solid. Such film with semi-solid properties cannot be further reworked without leaving permanent defects. Bearing that in mind we define this characteristic time point as the upper bound for the paint film workability and adopt it as a standard measure for the open time  $t_{OT}$ in the remainder of this paper. 
\subsection{Open time as function of film thickness}
We have have established how to quantitatively determine the open time of waterborne paints deploying LSI. Next, we will try to couple the open time to a related physical phenomenon with an already established analytical model. Being successful in such endeavor will allow us to: i) validate our LSI findings for the open time; ii) gain some predictive power that would enable us to forecast the open time as a function of different parameters that we deem relevant for the drying process. Since solvent evaporation is the driving force of the drying process, it is logical to use Langmuir's equation (LE) for the evaporation rate as s staring point:\cite{Langmuir1932}
\begin{equation}
\label{eq5}
\frac{\mathrm{d}M}{\mathrm{d}t} \, =(p_s-p_p) \sqrt{\frac{m}{2{\pi}kT}}
\end{equation}
Here $dM/dt$ is the evaporation rate (expressed in the net mass evaporating per unit time and unit area), $p_s$ and $p_p$ are the equilibrium and partial vapor pressure of water, $m$ is the mass of the water molecules, $k$ is Bolmetzmann's constant and $T$ is temperature.This expression is a convenient starting point because the mass change of the drying film is temporally followed in parallel with the LSI measurements (Fig. \ref{fig3}a) and by that we have direct access to the evaporation rate $dM/dt$. As a first simple case we will link $t_{OT}$ to the starting thickness $h_S$ of the wet paint film. LE instructs that the evaporation rate is independent of the starting thickness which is experimentally confirmed by LSI measurements (Fig. \ref{fig3}b). For convenience LE can be multiplied  by the surface area of the paint film $S_f$ to get the the total mass flux per unit time $Q_t=dM_t/dt$ which is a quantity directly accessible from the mass measurements. Then, after some rearrangement and subsequent integration we get:
\begin{equation}
\label{eq6}
\int_{M_S}^{M_{OT}}dM_t=\int_{t_S=0}^{t_{OT}}S_f(p_s-p_p) \sqrt{\frac{m}{2{\pi}kT}}dt
\end{equation}
\begin{equation}
\label{eq7}
M_S-M_{OT}=cS_ft_{OT}
\end{equation}
where $M_S$ is the starting mass of the film, $M_{OT}$ is the mass at $t=t_{OT}$ and $c$ is an aggregate of all remaining constants. We expect that the end of the open time period occurs always at the same effective solid contents despite of the starting thickness of the wet film. Thus, the relative mass $\phi_{OT}=M_{OT}/M_S$ at $t_{OT}$ remains constant for all film thicknesses. To test the validity of this conjecture, we have have computed $\phi_{OT}$ for paint films with different $h_S$. The experimentally determined $\phi_{OT}$ indeed seems to retain a constant value for all of the experimental conditions (Fig. \ref{fig3}c). We can then rewrite the mass difference  $M_S-M_{OT}$ as $M_S(1-\phi_{OT})$. The dimensions of paint film are imposed by the used deposition method (see 
Supporting information) and its width $w$ and length $l$ are the same for all experiments. Since the mass of the paint film is simply $M=wlh\rho$, eq. \ref{eq7} could be rewritten as:
\begin{equation}
\label{eq8}
t_{OT}=h_S\frac{\rho_{OT}(1-\phi_{OT})}{c}
\end{equation}
By omitting all the constants we get the scaling relation $t_{OP}\sim h_S$ which casts a linear dependence on the open time with respect to the staring thickness. This prediction is fully in line with the experimental data (Fig. \ref{fig3}d). The open time period is indeed prolonged in a linear fashion with respect to the increasing starting thickness of the paint film. The good agreement between experimental findings and prediction demonstrates the ability of this approach to quantitatively couple the experimentally determined open time with the underlying drying process. That allows us to use the same strategy in attempt to connect important environmental parameters with the open time of the drying paint film.     

\subsection{Effect of relative humidity on the drying dynamics and the open time}
Humidity has a profound effect on the drying dynamics of aqueous solutions, hence strongly influences the open time of waterborne paints. To capture this effect in a quantitative manner by deploying the newly developed LSI based approach, the relative humidity $H_R$ needs to be incorporate as a parameter in LE. That could be achieved by simply rewriting the later as:
\begin{equation}
\label{eq9}
Q_t =(a-H_R)p_sS_f \sqrt{\frac{m}{2{\pi}kT}}
\end{equation}
where  $a$ is a correction factor for the deviation of the water activity from the ideal values. After integration and some algebra we arrive at:
\begin{equation}
\label{eq10}
t_{OT}=\frac{c_1}{(a-H_R)}
\end{equation}
\begin{equation}
\label{eq11}
c_1=\frac{(M_S-M_{OT})}{p_sS_f}\sqrt{\frac{2\pi kT}{m}}
\end{equation}
Since $c_1=const$, the open time should scale with the inverse of the term $(a-H_R)$, $t_{OT}\sim1/(a-H_R)$. To test this prediction we first examine the temporal evolution of $\alpha$ for all of the $R_H$ levels used (Fig. \ref{fig4}a).
From the characteristic kinks in $\alpha(t)$ - curves we can easily estimate   
 the $t_{OT}$ for the different experimental conditions. Overlaying the experimentally determined $t_{OT}(H_R)$ and the prediction according eq.\ref{eq10} reveals an interesting finding. The theoretical prediction holds for the upper range of relative humidities $H_R\geq0.5$ but fails to describe accurately the experimental data in the lower range $H_R<0.5$ (Fig. \ref{fig4}b). Notably, there is an abrupt drop in the open time at $H_R\approx0.48$ and  barely any changes in its values after that. This discontinuity in the $t_{OH} (H_R)$ hints the advent of an additional phenomenon that alters the drying dynamics and by that the behavior of the open time in more dry conditions. Lower relative humidities enhance the drying kinetics which in turn could facilitate particle packing at $\phi_{OT}$ different from the one observed for more humid air or the formation of a thin solid layer at the film-air interface, a phenomenon also known as skin formation.\cite{Marin2011,Piroird2016,Noirjean,Sheetz1965,Routh2004,Wahdat2018} The occurrence of both phenomena could account for the sudden drop and the subsequent stationary value of the open time at lower $H_R$. To pinpoint the culprit for these experimental observations we first need to identify a characteristic signature for each of the proposed phenomena. If skin formation is taking place at $H_R<0.5$ then the mass transport through the air-film interface should be significantly impeded by the presence of a solid layer there. Thus, we would expect the change in the evaporation rate $Q_t$ to exhibit different regimes at lower and higher humidities. On the other hand, a different packing at  $\phi_{OT}$  should not interfere with the solvent evaporation. The length scale of the voids between the colloid particles in a close packing configuration is still much bigger than the size of the water molecule. Thus, the solvent transports towards the air-film interface should not be hindered and the change in $Q_t$ should follow the same trend throughout the full range of experimental conditions used. Eq. \ref{eq9} instructs us that the evaporation change should scale linearly with the relative tumidity: $Q_t\sim H_R$. This is indeed what we observe experimentally for $H_R\geq0.5$ (Fig. \ref{fig4}c). However, for $H_R<0.5$ the change in  the experimentally determined rate of evaporation significantly deviates from the linear regime that is exhibited at high humidities. The value of $Q_t$ barely changes and remains close to constant (Fig. \ref{fig4}c). Such behavior is inline with the scenario of skin formation. To further asses this scenario as a plausible explanation for the observed behavior of $t_{OT}(H_R)$, we have estimated the initial Peclet number $Pe_i$ for the paint films in the beginning of the drying process (see Supporting Information). The $Pe_i$ values are bigger than unity for  the whole $H_R$ - range implying that the system is predisposed to skin formation (Fig. \ref{fig4}d). Nevertheless, our LSI measurements indicate that skin formation is taking place only at $H_R<0.5$. Most likely, the disjoining pressure is overcome and coalescence in the layer of accumulated particles occurs only at such drying conditions (see Fig. S1 and  the corresponding discussion in Supporting Information).\cite{Feng} Both findings, the non-uniform behavior of evaporation rate as function of relative humidity and the high Peclet number values, strongly suggest that skin formation is most probably the cause for the discrepancy between the measured and forecasted $t_{OT}$ by eq. \ref{eq10}. 
\section{Open time versus temperature.}
Temperature is the other crucial environmental parameter that has a profound effect on the drying in aqueous systems. To quantify the effect of temperature on the open time of drying paint films, we need to incorporate all its contributions in LE. Considering solely the original form of LE (eq. \ref{eq5}) leads to the relative simple scaling $t_{OT}\sim \sqrt{T}$. However, $p_s$ is strongly dependent on temperature and this needs to be accounted for. The well known Clausius-Clapeyron equation (CCE) is an ideal gas based approximation that gives the relation between temperature and equilibrium vapor pressure $p_s$:\cite{Clausius1850}
\begin{equation}
\label{eq12}
\ln{p_s}=-\frac{L}{RT}+C
\end{equation}
Here $L$ and $R$  are the specific enthalpy of vaporization and the universal gas constant respectively. CCE assumes a temperature independent $L$ which is a good approximation for low temperatures. At standard atmospheric conditions however, the temperature dependence of $L$ (and $C$) cannot be neglected. To that end we invoke the August-Roche-Magnus equation which is a good empirical approximation widely used in atmospheric science:\cite{Alduchov1996} 
\begin{equation}
\label{eq13}
p_s=610.94e^{\frac{12.27T}{T+273.3}}
\end{equation}
Substituting eq. \ref{eq13} in eq.\ref{eq5} and omitting the constants yields a non-trivial relation between temperature and the open time:
\begin{equation}
\label{eq14}
t_{OT}\sim\frac{\sqrt{T}}{le^{\frac{mT}{T+n}}}
\end{equation}
where $l$,$m$ and $n$ are the empirical constants from eq. \ref{eq13}. The the experimental data for $t_{OT}$ at different temperatures (Fig. \ref{fig5}) is in unison with scaling relation \ref{eq14}.   

\section{Beyond the open time}

We have shown that LSI allow us to closely follow the dynamics in the early stages of drying colloidal films and by that to determine accurately and unambiguously an important quantity such as the open time.  However, other phenomena that populate the later drying stages are of equal importance for the terminal state of the paint film. The logical question that arises is if LSI can be also used to monitor these later events. The film formation stage is determinant for the appearance and performance of the dried coating.  At the heart of the film formation is the process of phase inversion in which the  binder particles coalesce and give rise to a continuous polymer phase with the residual water suspended in it as droplets. After the binder particles attain close packing and undergo deformation, coalescence commences with the breakage of the thin water film that disjoins the highly deformed particles. The rupture of the water film is caused by the capillary forces which continuously grow with the ongoing solvent evaporation and at a certain point overwhelm the disjoining pressure keeping the soft colloids apart. Once the film fails, the contents of the particles that have been kept apart by it intermix. {\it In situ} methods to study the  process of phase inversion in a spatially resolved manner are scarce. It is challenging to detect coalescence events by simply watching for structural changes in the system under investigation, since  there is no clear macroscopic difference between strongly deformed and coalesced particles. In some cases, microscopy approaches could be deployed to monitor particle deformation and coalescence but their application becomes limited if the studied system is opaque and once the size of the coalescing particles approaches the diffraction limit.\cite{Mchugh1992,Charin2013,Gonzalez2012, Krebs2013}. To circumvent these challenges we use LSI  as an approach that relies on dynamics to generate contrast and follow in details the film formation process in opaque drying paint films. 
We start by mapping the dynamics in drying films of acrylic emulsions with different concentrations of coalescing agents (see Experimental section). The d2 maps for a film containing 2,7 wt$\%$ coalescent reveal a sudden enhancement in the dynamics much later than the expiry of the open time period  (Fig. \ref{fig6}a, b). This enhancement is followed by the passage of a well defined front of even more intensified dynamics (Fig. \ref{fig6}c) and a subsequent attenuation (Fig. \ref{fig6}d). To dissect this sequence of events in a more quantitative manner we compute the temporal evolution of the field correlation function $g_1(\tau)$ and then extract the relaxation time $\tau_0$ and stretching exponent $\alpha$ for different time points using eq.\ref{eq. 4}. The $g_1(\tau)$  undergoes a two orders of magnitude shift to much larger $\tau$-values indicating a significant dynamics retardation in the period succeeding the open time point (Fig. \ref{fig6}e). This retardation trend is interrupted by a temporary acceleration in the dynamics which is accompanied also by an alteration in the shape of the  $g_1(\tau)$-curve signaling for a change in the particles mode of motion.
Close examination of the of the $\tau_0$-curves allow us to identify four distinct drying stages for all coalescent concentrations (Fig. \ref{fig6}f). Stage \textrm{I} is essentially the open time period of the film which was extensively discussed already. Stage \textrm{II} is marked by transient decrease of $\tau_0$, implying enhanced motion. Most likely the rising capillary pressure compresses and deforms the soft binder particles. This deformation translates to a macroscopic contraction of the film imposing a net translation motion of the scattering TiO\textsubscript{2} particles downwards and towards the center of the drying film. Once the capillary pressure exceed the disjoining pressure, the drying process commences with stage \textrm{III} in which the interstitial water film ruptures and phase inversion takes place. Once the binder particles coalesce, the TiO\textsubscript{2} particles suddenly find themselves in a continuous polymer phase with a significantly higher viscosity which is reflected by the steep rise in $\tau_0$. Additionally, the accelerating ballistic motion of the particles shifts to a more diffusive one. The onset of the steep rise in the $\tau_0$-curve marks the time point to which we refer as the film formation time $t_{FFT}$. In the final stage \textrm{IV},  the coalescing aids slowly evaporate from the film, causing the viscosity to gradually increase and the mobility of the TiO\textsubscript{2} particles to drop further. The resultant rise in $\tau_0$ continues over many hours to days (Fig. \ref{fig6}f, inset).\newline
The temporal behavior of $\tau_0$ and $d_2$ maps fall well inline with each other. The rapid increase in $\tau_0$ implies that coalescence must occur across a sharp front, as was found previously.\cite{Ma2005,Haley2008,Vanderkooij2015}
This is indeed what we have clearly observed in the $d_2$ maps for the film containing 2.7 wt\% coalescent at $t = t_{FFT}$ (Fig.~\ref{fig6}f). A well defined gradient in mobility (Fig.~\ref{fig6}c) traverses through the field of view (Supplementary Video). This front is truly a dynamic transition and cannot be identified based on structural features, i.e.~relying on absolute intensity (Fig. S3a). Our interpretations are further supported by the $\alpha(t)$ curves. The significant increase of $\alpha$ in stage \textrm{II} implies for the onset of  ballistic dynamics (Fig. \ref{fig6}h) which is in line with the stipulated collective translation of the particles caused by the contracting coating. Additionally, in stage \textrm{III} we observe a rapid drop in $\alpha$ after the film formation time (Fig.~\ref{fig6}h). This sharp decrease in $\alpha$ combined with the kink in $\tau_0$ is in agreement with our hypothesis that across the front, the TiO\textsubscript{2} particles shift from accelerating translation to more diffusive dynamics which may well be the signature of coalescence in a coating.\newline
Even though the overall shapes of the $\tau_0$ and $\alpha$ curves are the same for different coalescent concentrations, they are clearly shifted along the time axis (Fig.~\ref{fig6}f, h). Interestingly, adding coalescence agents considerably decreases the open time (inset Fig. \ref{fig6}); this is an undesired effect which is important to take into account when formulating paints. We attribute this effect to the partition of coalescing aids into the polymer particles. As a consequence the binder particles swell and close packing is reached at an earlier time point. Dynamic light scattering measurements confirm this difference in particle sizes: the samples with 2.7 and 7.1 wt\% coalescent exhibit particle diameters of 124 and 142 nm, respectively.

Fortunately, not only the open time decreases upon increasing the coalescent concentration, the film formation time decreases even more, which is the intended result. Coalescing aids are commonly added to plasticize the binder particles by increasing the free volume of the polymer chains.\cite{Schroeder2010,Divry2016,Berce2017} The earlier onset of coalescence is further enhanced by the swelling-induced increase in the volume fraction of the latex particles. Another consequence is the absence of a sharp coalescence front in the strongly plasticized paint film (7.1 wt\% coalescent), which instead is much more gradual and delocalized (Fig. S3c). These observations are in agreement with previous work on drying surfactant-stabilized oil-in-water emulsions, which revealed two distinct modes of coalescence: front coalescence at high surfactant concentrations (stable emulsions) versus bulk coalescence at low surfactant concentrations (unstable emulsions).\cite{Feng} In the first scenario, a steep gradient in the capillary pressure is required to induce coalescence. The later takes place at only at the drying end of the emulsion where the high capillary pressure condition is met. In the second scenario, a lower capillary pressure is already sufficient to induce coalescence, which thus occurs simultaneously at many locations in the bulk. These results are analogous to the findings in our work, where instead we have varied the particle resistance to coalescence by changing the coalescent concentration.\par

Since the characteristic shape of $\tau_0(t)$ and the dip in $\alpha (t)$ are reproducible in all measurements and for other latex systems, they may serve as standard approach to extract the film formation time in drying waterborne coatings. The small standard deviations in $t_\mathrm{FFT}$ (Fig.~\ref{fig6}g, inset) which were derived from the onset of the abrupt decrease in $\alpha$ highlight the robustness of this procedure. Although the observed strong dependence of coalescence on the coalescent concentration is in line with expectations and earlier research, we now have a straightforward and easily adaptable method to quantify the film formation time unambiguously as an addition to the LSI based approach for open time determination that was demonstrated earlier. \cite{Schroeder2010,Divry2016,Berce2017}

\section{Conclusions}

In this work we have introduced Laser Speckle Imaging in the light of a versatile tool for a quantitative visualization of drying waterborne paint films. As such, LSI enables us to unambiguously determine the open time of realistic paint formulations and for wide range of drying  conditions. The good agreement between our experimental observations for the open time and the scaling predictions derived from Langmuir's evaporation rate theory not only validates our LSI approach  for open time determination but also allows accurate predictions for the behavior of this quantity in the whole set of conditions used in the current work. Furthermore, the advent of deviations from the expected behavior permits the detection of other important phenomenon that might occur during drying such as skin formation. The wide temporal dynamic range of LSI  makes possible the extension of our quantitative observations towards a bigger timescale that is populated by another set of relevant phenomena, namely the deformation of binder particle in later stages of the drying process and the subsequent coalescence into a solid paint film. LSI allow us not only to accurately determine the time point in which phase inversion sets in but also to monitor the propagation of the coalescence front in a spatio-temporal resolved manner. Summarizing all the findings in this work convincingly outlines LSI as a valuable assistant in paving the way for the development of water-based alternatives of VOCs rich coatings. 
 
 \newpage
\section{Figures}
%
\begin{figure}[H]
\hspace{-14 mm}
  \includegraphics[width=180 mm]{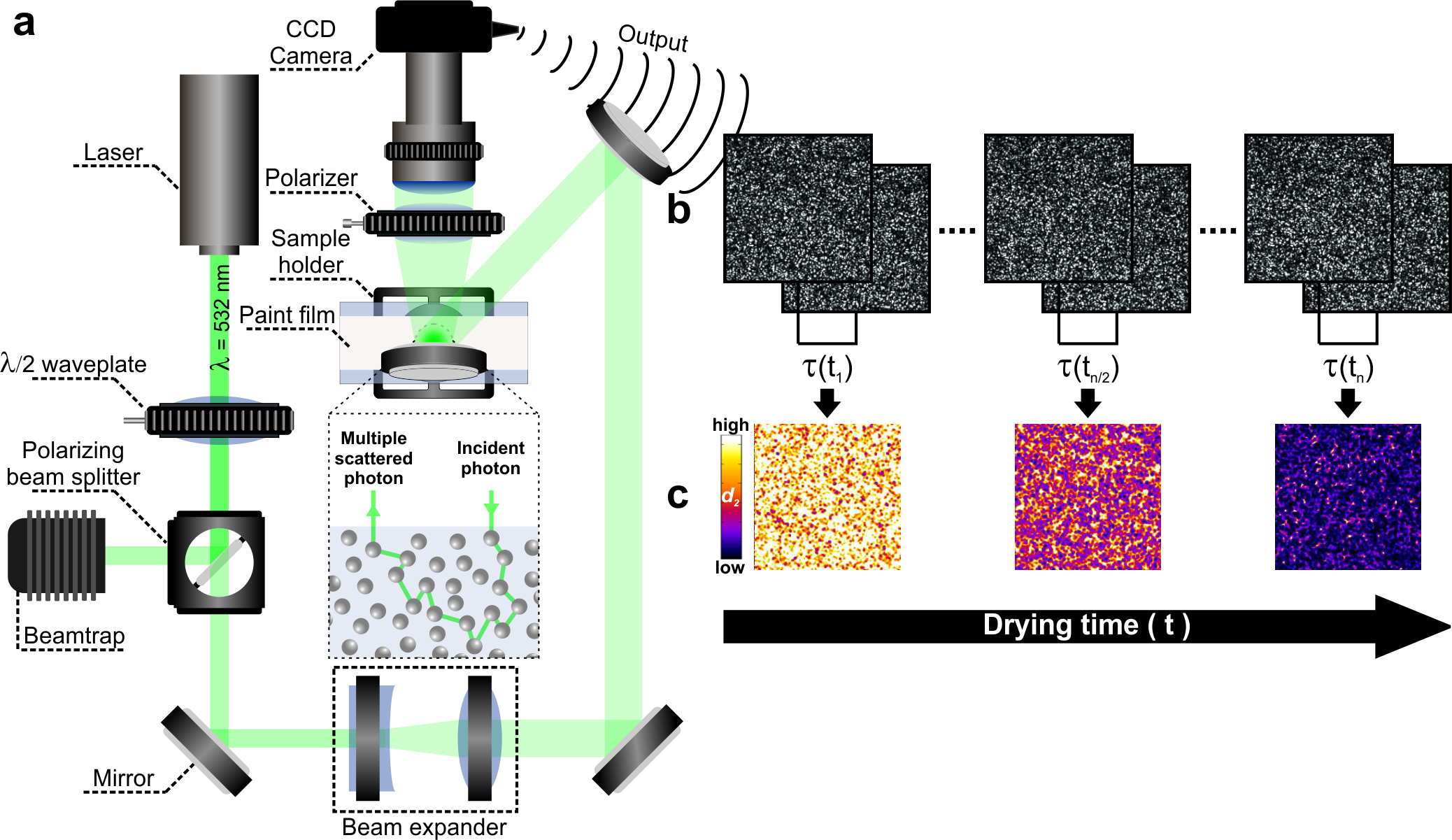}
  \caption {\textbf{LSI setup and output} (\textbf{a}) Scheme of the LSI setup. The light is guided to the drying aqueous paint film and impinges it. After multiple scattering events the propagation direction of photons in the sample is completely randomized (inset image in \textbf{a}).  The multiple scattered light emanating from the sample is then collected and projected on a CCD chip. This yields a sequence recording of speckle pattern images (\textbf{b}). Correlating the raw speckle images using eq.\ref{eq. 1} and a fixed correlation time $\tau$ at different time periods $t_i$ ($i=1..n$) of the drying process translates the raw data into dynamic maps $d_2$  (\textbf{c}). In the beginning $t_1$ of drying process the film is populated with high dynamics while at the mid-drying stage $t_{n/2}$ the dynamics are significantly slowed down due to the ongoing evaporation. At end $t_n$ all the dynamics on the timescale of $\tau$ are brought to a halt.}
  \label{fig1}
\end{figure}
\newpage
\begin{figure}[H]
\vspace{4mm}
  \includegraphics[width=120mm]{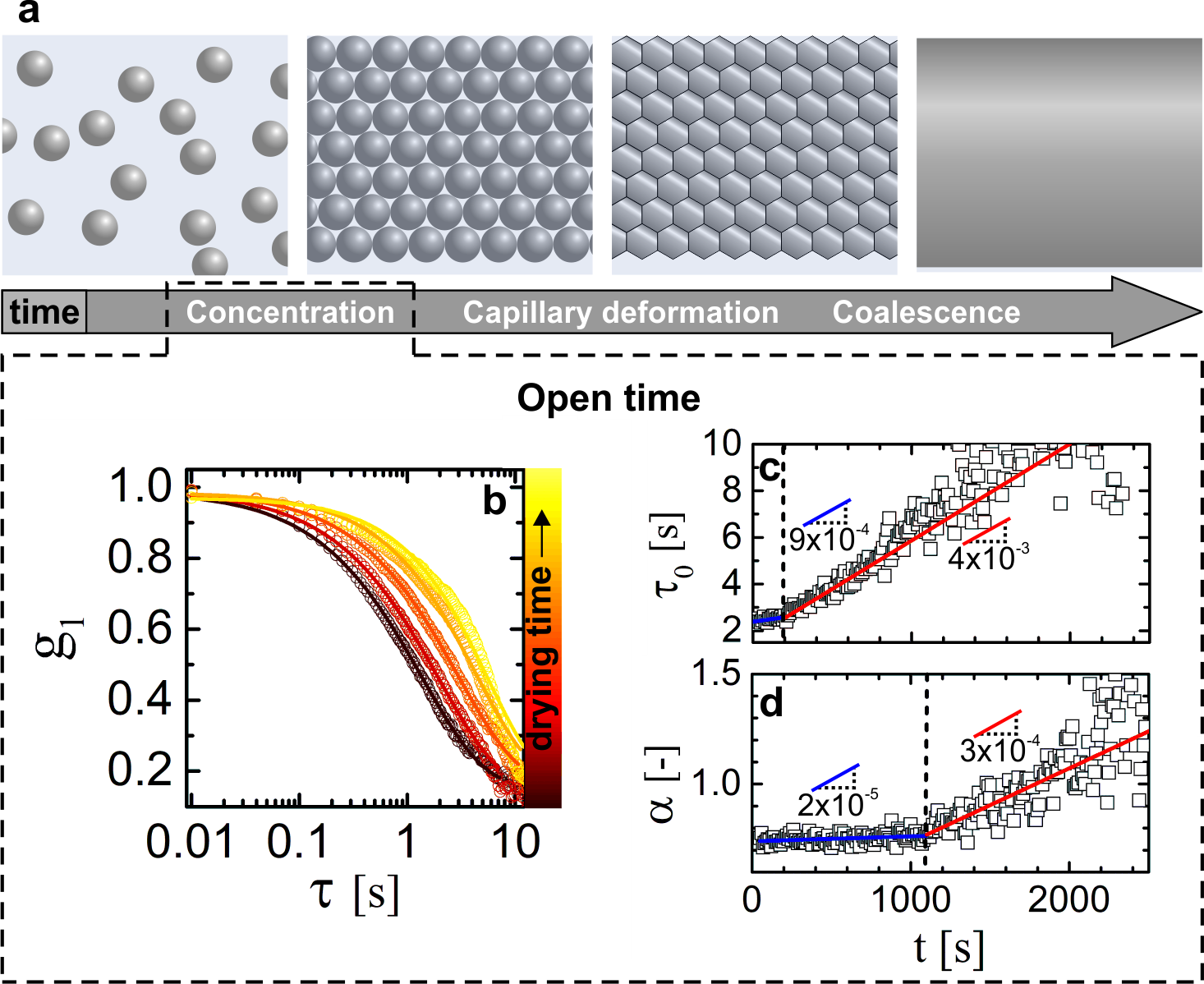}
  \caption{\textbf{Open time determination with LSI.} (a) Simplified schematic of the main events taking place during the drying of aqueous film of soft colloidal particles. (b) Temporal evolution of the field correlation function $g_1$. The curves shift towards longer correlation times $\tau$ due to a slow down in the dynamics caused by the solvent evaporation. (c) Changes in the relaxation time $\tau_0$ and (d) the stretching exponent $\alpha$ with the progression of the drying process. The temporal evolution of both parameters could be easily divided in two distinct regimes demarcated with a dashed line and colored (solid lines) in blue or red. The slopes (\raisebox{+1pt}{\fbox{\protect\includegraphics[height=9pt]  {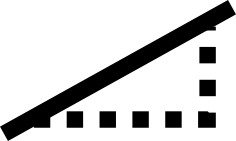}}}) $d\tau / dt$ and $d\alpha / dt$ are colored in accordance with the corresponding regime. For $\tau_0(t)$ the dashed line marks the start of a sudden and more steep increase of the viscosity. This time point marks the onset of deterioration in the workability of the film. In the case of $\alpha(t)$ the dashed line designates the abrupt change in the stretching exponent which signals the formation of a semi-solid film. This time point is viewed as the upper bound of the open time for the drying film.} 
  \label{fig2}
\end{figure}
\newpage
\begin{figure}[H]
\vspace{50mm}
  \includegraphics[width=175mm]{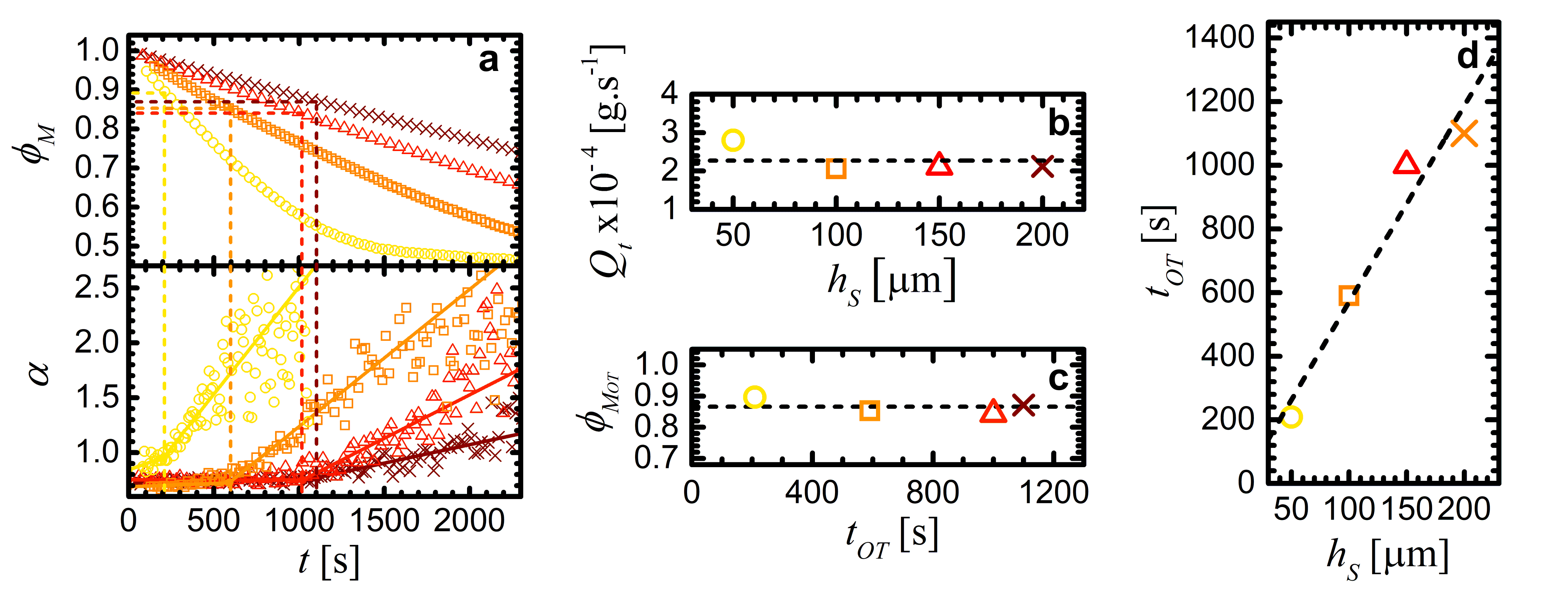}
  \caption{\textbf{Open time as function of film thickness.} (a) Mass fraction $\phi_M$ (top) and stretching exponent $\alpha$ (bottom) as function of drying time for films with thickness of 50 $\mu m$   (\raisebox{-.5ex}{$\textbf{\textcolor{mycol1}\SmallCircle}$}), 100 $\mu m$   (\raisebox{-.5ex}{$\textbf{\textcolor{mycol2}\SmallSquare}$}), 150 $\mu m$  (\raisebox{-.5ex}{$\textbf{\textcolor{mycol3}\SmallTriangleUp}$}) and
200 $\mu m$ (\raisebox{-.5ex}{$\textbf{\textcolor{mycol4}\SmallCross}$}). The dashed lines show the mass fractions at the end of the open time period $\phi_{M_{OT}}$  for the films with different thickness $h_S$. The solid lines guide the eye through the different regimes in the $\alpha(t)$ - curves and help to designate the open time point $t_{OT}$. The total mass flux per unit time $Q_t$ (b) and the mass fraction $\phi_{M_{OT}}$ (c) remain constant for all film thicknesses. Symbols represent the experimental data while the dashed lines guide the eye through the data points. (c) Open time as function of film thickness. The open time determined with LSI scales linearly with $h_S$ as predicted by eq. \ref{eq8} (dashed line).}
  \label{fig3}
\end{figure}
\newpage
\begin{figure}[H]
  \includegraphics[width=85mm]{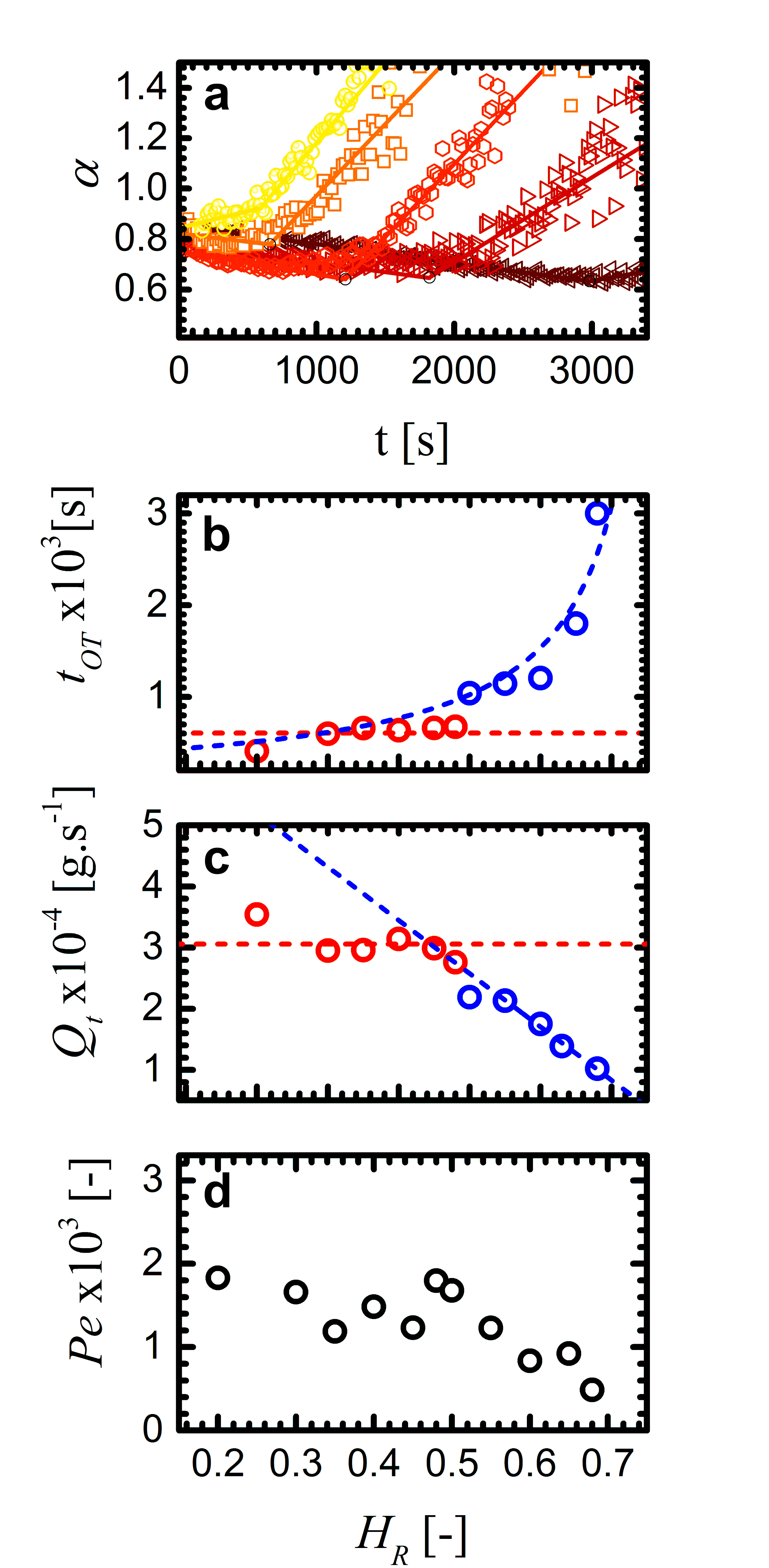}
  \caption{\textbf{Open time as function of relative humidity.} (a) Temporal evolution of the stretching exponent $\alpha$ for 200 $\mu m$ films at  relative humidities of $H_R=0.3$ (\raisebox{-.5ex}{$\textbf{\textcolor{mycol1}\SmallCircle}$}), $H_R=0.45$ (\raisebox{-.5ex}{$\textbf{\textcolor{mycol2}\SmallSquare}$}), $H_R=0.50$ ({$\textbf{\textcolor{mycol3}\hexagon}$}), $H_R=0.60$ (\raisebox{-.5ex}{$\textbf{\textcolor{mycol6}\SmallTriangleRight}$}) and $H_R=0.68$ (\raisebox{-.5ex}{$\textbf{\textcolor{mycol4}\SmallTriangleLeft}$}). For clarity, just a part of the experimental data is included. (b) Open time as function of relative humidity. The experimental data obtained for $R_H\geq0.5$ (\raisebox{-.5ex}{$\textbf{\textcolor{mycol7}\SmallCircle}$}) is in good agreement with the prediction according eq. \ref{eq10} (blue dashes). Due to skin formation the open time $t_{OT}$ exhibits a sharp drop and barely any changes for $R_H<0.5$ (\raisebox{-.5ex}{$\textbf{\textcolor{red}\SmallCircle}$}). (c) Evaporation rate vs relative humidity. The evaporation rate (\raisebox{-.5ex}{$\textbf{\textcolor{mycol7}\SmallCircle}$}) behaves in accordance with the expected scaling  $dM_t/dt\sim R_H$  (blue dashes) for $R_H\geq0.5$ but attains a constant value at $R_H<0.5$ (\raisebox{-.5ex}{$\textbf{\textcolor{red}\SmallCircle}$}), suggestive for skin formation. (d) Peclet number vs relative humidity.}  
  \label{fig4}
\end{figure}
\newpage
\begin{figure}[H]
 \vspace{50mm}
  \includegraphics[width=85mm]{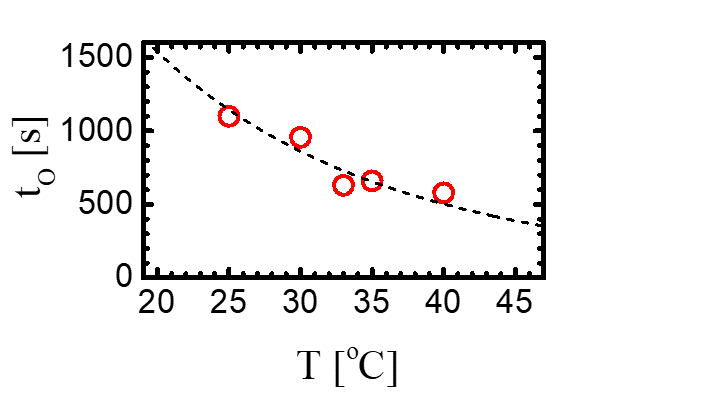}
  \caption{\textbf{Open time as function of temperature.} The symbols represent the experimental data while the dashed line shows the prediction according eq.\ref{eq14}.}
  \label{fig5}
\end{figure}
\newpage

\newpage
\begin{figure}[H]
\hspace{-20mm}
\includegraphics[width=160mm]{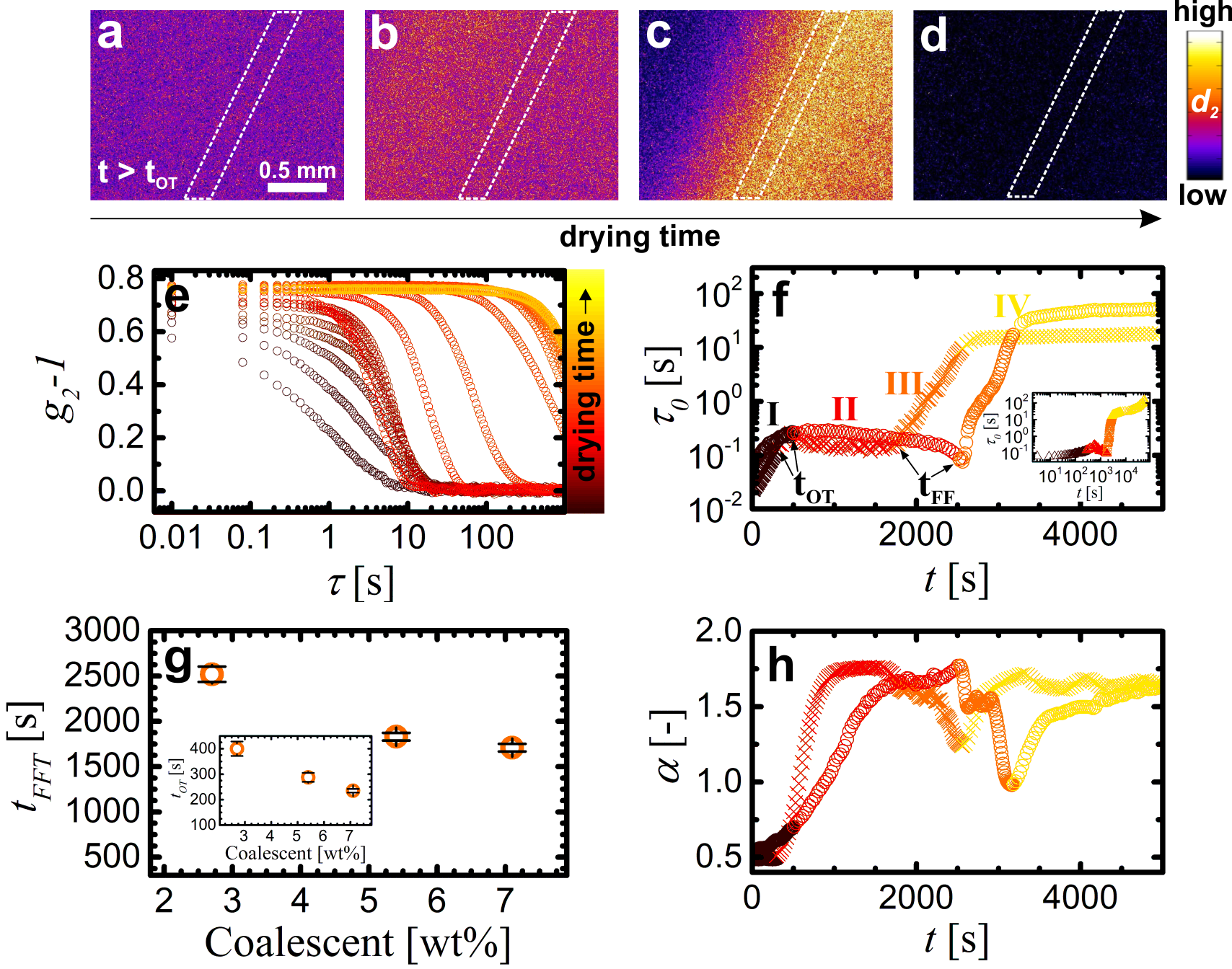}
\caption{\textbf{Quantification of film formation time and its dependence on coalescent concentration.} (a-d) $d_2$ maps of a 100 $\muup$m (wet-thickness) paint film at different time points beyond the open time and with 2.7 wt$\%$ coalescent concentration, capturing the dynamics accompanying the film formation process. At $t=t_{FFT} = 2550$ s the passage of a coalescence front (c) could be clearly identified. The maps were computed for correlation time $\tau = 2$ s. The dashed lines delineate the region over which the correlation curves (e), $\tau_0(t)$ (f) and $\alpha(t)$ (h) were computed. Similar regions of interest underlie all results in this figure.
(e) Typical correlation functions over time for the film containing 2.7 wt$\%$ coalescent showing a distinct crossover and change of shape upon film formation. The time interval between consecutive curves is 5 min.
(f) Temporal evolution of the characteristic relaxation time $\tau_0$ for 2.7 wt\% (\raisebox{-.5ex}{$\textbf{\textcolor{black}\SmallCircle}$}) and 7.1 wt\% (\raisebox{-.5ex}{$\textbf{\textcolor{black}\SmallCross}$}) coalescent concentration. The Roman numbers and corresponding colors designate the different stages in drying process as described in the main text. The inset in (f) shows a long term measurement of $\tau_0$ for the film with 5.4 wt\% coalescent. (g) Film formation time and open time (inset) as a function of coalescent concentration. Each data point is the average of two measurements; the error bars represent the standard deviations. (h) Temporal evolution of the stretching exponent $\alpha$.}
\label{fig6}
\end{figure}
\newpage
\section{Experimental}
\subsection{Experimental setup.}
All the measurements were performed on a custom built setup which is represented schematically in Fig. \ref{fig1}a. The sample holder is situated in an insulated chamber in which the temperature and relative humidity are maintained constant at the desired values. if note mentioned otherwise, the measurements are performed at room temperature and relative humidity of 50 \%.  The holder is coupled to a computer-controlled balance (WZA224-NC, Sartorius) which allows for continuous mass monitoring of the sample. A 532 nm solid state laser (Samba 1W, Cobolt) is used as a light source. The beam is first driven through a half-wave plate and then through a polarizing beam splitter in order to adjust the intensity. Subsequently, the light is guided through a beam expander to set the cross-section size of the beam  to $d\sim 1 cm$ . The resulting beam is then navigated with a system of mirrors towards the sample. The back-scattered light first passes through a polarizer which filters the reflected fraction and the photons with short paths emanating form the sample. Finally, the filtered back-scattered light is collected with a zoom lens (1.8x) and is focused on a CCD camera (Dalsa Genie, Stemmer Imaging). The speckle size is set 2-3 times bigger than the physical size of the camera pixels to ensure a good balance between spatial resolution and signal-to-noise ratio.\cite{Kooij2016} All the LSI images are acquired at 100 fps with exposure optimized to exploit the full dynamic range of the CCD camera. 

\subsection{Data analysis.}
To study the ongoing dynamics in the drying samples we calculate the intensity structure factor $d2$, the intensity correlation function $g_2$ and the field correlation function $g_1$ using eq.\ref{eq. 1}, eq.\ref{eq. 2} and eq.\ref{eq. 3} respectively. The spatial correlation factor $\beta$, which accounts for the number of speckles per camera pixel ($\beta<1$), is chosen such $g_2-1\rightarrow \beta$ for $\tau \rightarrow 0$. The value of 1.5 is used for the numerical prefector $\gamma$ which has been experimentally determined elsewhere.\cite{Kooij2016}

\subsection{Materials.}
All the measurements were performed on commercial water-based paints (acrylic styrene copolymer emulsions). The studies investigating the effect relative humidity ($T = 25^\circ C$) and temperature ($H_R=0.5$) were performed using 200 $\mu$m thick paint films deposited using a custom made film applicator ensuring a wet film with well defined dimensions (see Supporting Information). The paint formulation includes  2.7 wt\% butyldiglycol and texanol as coalescing aids, 0.15 wt\% TiO$_2$ (d = 316 $\pm$ 8 nm) as pigment and total solids of $\sim$ 40 wt\%. The film formation studies in the later drying stages were performed on 100 $\mu$m thick films of paints with varying total of 2.7 wt\%, 5.4 wt\% and 7.2 wt\%  butyldiglycol and texanol as coalescing aids. The temperature and relative humidity were fixed at 23 $\pm$ $1^\circ$C and 43 $\pm$ 2\% respectively. For all of the measurements, unless specified differently, the paint films were deposited on a glass substrate.

\begin{acknowledgement}

This research forms part of the research program of the Dutch Polymer Institute (DPI), projects 913ft16 and 781.

\end{acknowledgement}

\begin{suppinfo}

The following files are available free of charge.
\begin{itemize}
  \item opentimeSI.pdf: supplementary figures and discussions about the Peclet number determination, the paint deposition method and coalescence in the paint films.
  \item coalescence\_x30.avi: $d_2$ video capturing the coalescence in a paint film.
\end{itemize}

\end{suppinfo}

\section{SUPPLEMENTARY MATERIAL}
\subsection{Peclet number estimation and skin formation in drying paint films.}
To further asses the skin formation susceptibility of the sample we have estimated the initial Peclet number $Pe_i$ for the different measurement conditions. In a colloidal suspension with an evaporating continuous phase, there is an interplay between two types  types of mass transport : i) diffusive as a consequence of Brownian motion; and ii) advective directed upwards due to the mass flux generated by the ongoing evaporation at the air-water interface. The Peclet number estimates the balance between those two  modes of mass transport and reads:

\begin{equation}
\label{eqS1}
Pe_i=\frac{h_sv}{D}
\end{equation}

where $h_s$ is the starting thickness of the film, $v$ is the velocity with which the air-water interface of film moves moves down ($dh/dt$) and $D$ is the diffusion constant for scattering particles. We have access to all of these quantities which allow us to estimate $Pe_i$. The starting thickness $h_s$ is known and fixed by the methodology used to deposit the paint film. The velocity $v$ can be easily calculated from the temporal resolved gravimetric measurements of the drying film.  From the LSI measurements we can compute the relaxation time $\tau_0$ and calculate $D$ through the relation: $\tau_0=1/(6k_0^2D)$, where $k=2\pi n/\lambda$ is the wave vector.\par

For $Pe$ bellow or close to unity, the diffusive transport ensures a uniform particle distribution, and hence, uniform pressure throughout the sample (Fig. \ref{figS1}a, b). For $Pe>1$, particles accumulate at the air-water interface and a concentration gradient evolves which gradually decays away from the interface $h$ (Fig. \ref{figS1}c, d). High $Pe$ is necessary but not sufficient condition for skin formation to occur. In order for a solid film to form a critical particle fraction $\phi_c$ (pressure higher than the disjoining pressure) needs to be attained which will ensure the onset of coalescence (Fig. \ref{figS1}e, f). 

\begin{figure}[H]
\label{figS1}
\includegraphics[width=120mm]{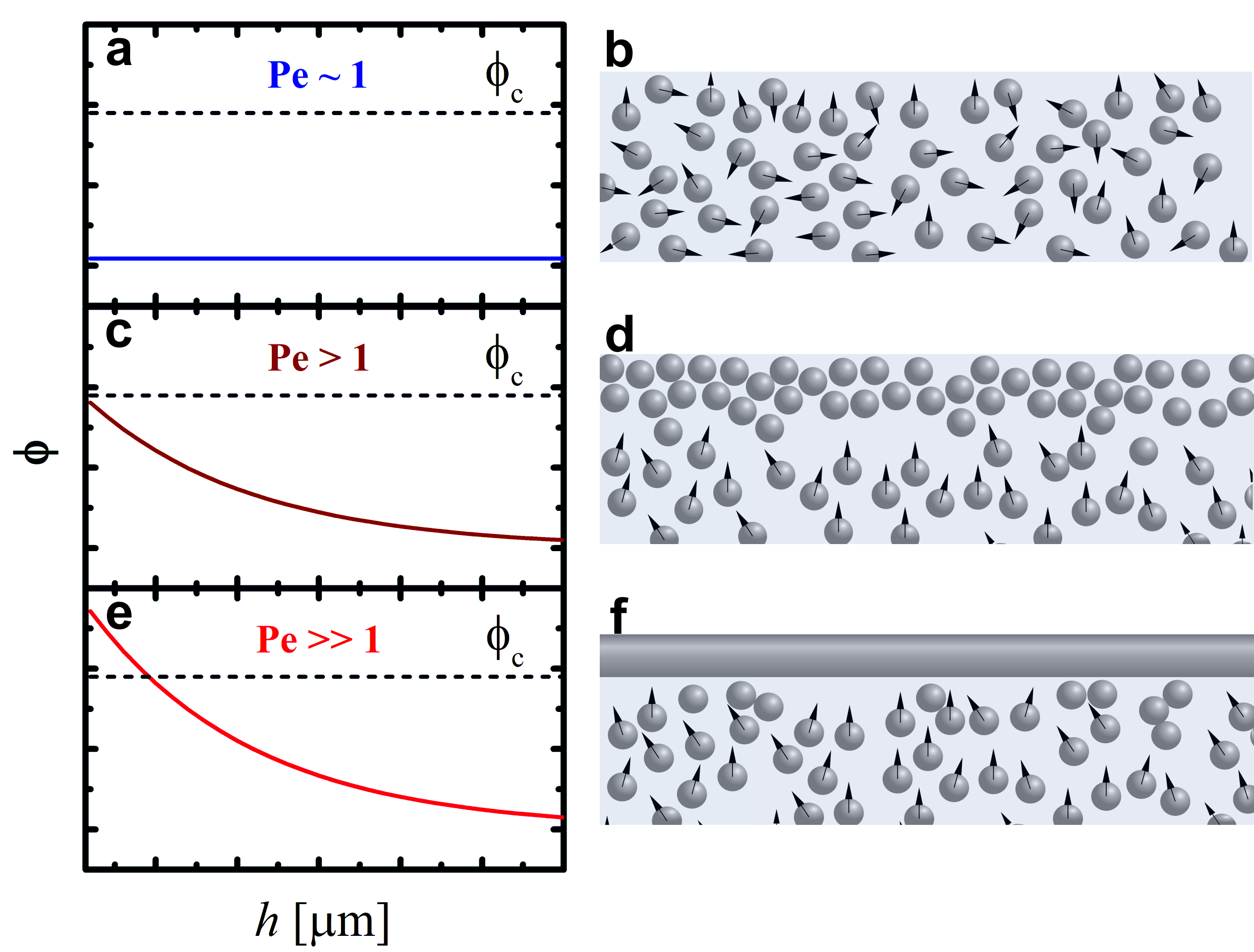}
\renewcommand{\thefigure}{S\arabic{figure}}
\caption{\label{figS1} Peclet number and skin formation in drying suspensions of soft particles. Tentative particle fraction $\phi$ profiles for suspensions drying at different $Pe$ regimes (a,c and e) and the corresponding schematics (b,d and f) representing the author's impression of the accompanying phenomena.}
\end{figure}

\newpage
\subsection {Film deposition method}
\begin{figure}
\includegraphics[width=100mm]{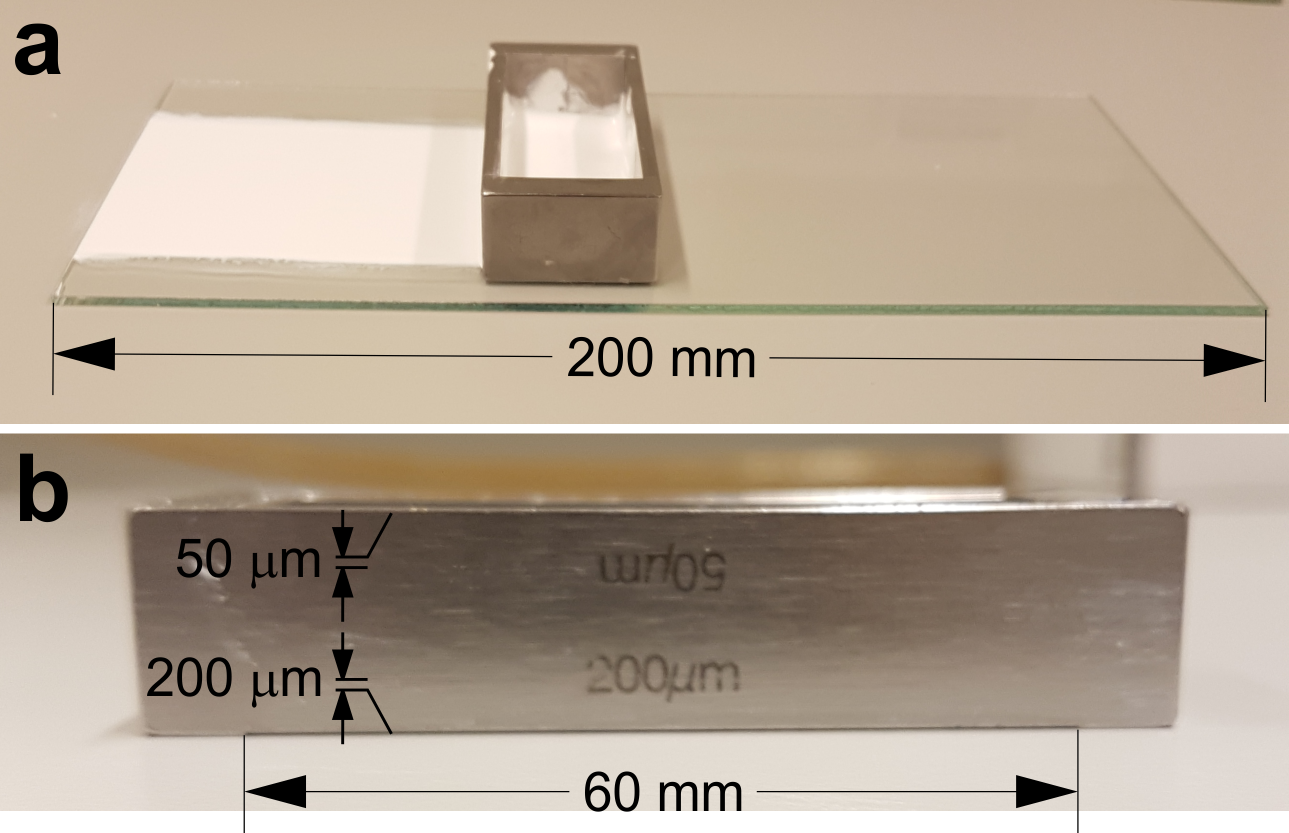}
\renewcommand{\thefigure}{S\arabic{figure}}
\caption{\label{figS2} Deposition method. (a) Deposition of an paint film with a film aplicator on a glass slide. (b) Close up of the film applicator.}
\end{figure}

.
We  apply a homogeneous film of acrylic water-based paint on a glass surface using a film applicator (caster) (Fig. \ref {figS2} a) that has some if its edges partially indented to an extent that ensures the casting of a film with the desired thickness (Fig. \ref {figS2} b). Once the film is deposited we introduce the sample in the setup and commence with the measurement. 

\newpage
\subsection{Coalescence front in paint films.}
\begin{figure}
\includegraphics[width=170mm]{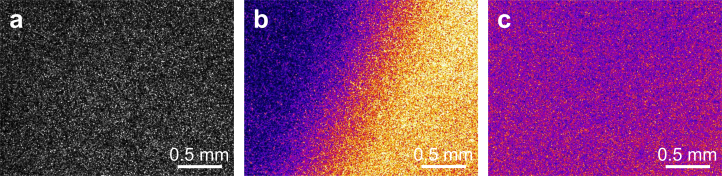}
\renewcommand{\thefigure}{S\arabic{figure}}
\caption{\label{figS3} Coalescence in waterborne latex films. (a) Raw speckle image and (b) a $d_2$ map of a coalescing paint film with 2.7 wt$\%$ coalescing aids. (c) $d_2$ map capturing the coalescence in plasticized paint film with 7.1 wt$\%$ coalescing agents.}
\end{figure}

Coalescence events cannot be detected simply by analyzing the light intensity in the captured raw speckle images (Fig. \ref{figS3}a) but can be clearly identified in the $d_2$ dynamic map of the drying paint film (Fig. \ref{figS3}b). The coalescence front is well defined  in the case of less soft binder particles in a paint film with 2.7 wt$\%$ coalescents (Fig. \ref{figS3}b). In contrast, the plasticized film with 7.1 wt$\%$ coalescents accommodates a more gradual and delocalized coalescence event (Fig. \ref{figS3}c). 


\end{document}